\documentclass[11pt]{article}
\usepackage{moriond,epsfig}

\def\Journal#1#2#3#4{{#1} {\bf #2}, #3 (#4)}

\def\NPB{{\em Nucl. Phys.} B}

\def\PRL{\em Phys. Rev. Lett.}

\def\JHEP{\em J. High Energy Phys.}
\def\CPC{\em Comput. Phys. Commun.}
\def\JPHGB{{\em J. Phys.} G}
\def\be{\begin{equation}}
\def\ee{\end{equation}}
\def\bea{\begin{eqnarray}}
\def\eea{\end{eqnarray}}

\begin{document}
\vspace*{4cm}
\title{NEXT-TO-LEADING ORDER $t\bar{t}$ PLUS JETS PHYSICS WITH HELAC-NLO}

\author{MA\L GORZATA WOREK \footnote{
Presented at the XLVth Rencontres de Moriond on QCD and High Energy 
Interactions, La Thuile (Aosta), Italy, March 13-20, 2010. }}

\address{Fachbereich C Physik, Bergische Universit\"at Wuppertal, 
D-42097 Wuppertal}

\maketitle\abstracts{   A report on the recent next-to-leading order QCD
  calculations to  $t\bar{t}b\bar{b}$ and $t\bar{t}jj$ processes  at the CERN
  Large Hardon Collider is presented. Results for integrated and differential
  cross sections are given. A significant reduction of the scale
  dependence is observed in both cases, which indicates that the perturbative
  expansion is well under control.  The results are obtained in the framework
  of the  \textsc{Helac-Nlo} system. }

\section{Introduction}

For a light Higgs boson, with mass $m_H \le 135$ GeV the highest decay rate 
mode is $H\to b\bar{b}$. In the dominant Higgs boson production 
channel at the CERN
 Large Hardon Collider (LHC), i.e. in gluon fusion, this decay mode is not 
very highly considered because of its overwhelming QCD background. On the 
other hand, the associated production of a top quark pair  with a Higgs boson
includes more distinctive signature, which should provide 
a unique opportunity of  independent  direct measurement 
of Higgs boson Yukawa couplings 
to tops and bottoms. Whether or not it will also provide a discovery channel, 
depends very much on the ratio between this signal and the main QCD 
backgrounds.  Early studies at ATLAS and CMS even suggested discovery 
potential, however, analyses with more realistic calculations of background 
processes, still only based on leading-order (LO) matrix elements, 
 show problems if the latter are not very well 
controlled \cite{Ball:2007zza,Aad:2009wy}. A careful and detailed examination 
of the backgrounds shows  that the most relevant are the direct production of 
the final state $t\bar tb\bar b$ (irreducible background) and the production 
of a top quark pair in association with two jets, $t\bar t jj$ 
(reducible background). The latter needs
to be taken into account due to the finite efficiency in identifying
b-quarks in jets (b-tagging). The calculation of the next-to-leading-order 
(NLO) QCD
corrections to  both background processes can be regarded as a major 
step forward towards the observability of the 
$t\bar{t}H \to t\bar{t}b\bar{b}$ signal at the LHC.  

The calculations of the NLO QCD corrections to the $pp\to t\bar{t}b\bar{b}$
process have been first presented last year 
\cite{Bredenstein:2009aj} and subsequently confirmed by our group at the
permille level \cite{Bevilacqua:2009zn}.  Very recently, the NLO QCD
corrections to $p p \to t \bar t jj$ background  process have been performed 
\cite{Bevilacqua:2010ve}. In this contribution, a brief 
report on these computations is given.

\section{Details of the next-to-leading order calculation}
 
The next-lo-leading order  
results are obtained in the framework of \textsc{Helac-Nlo} based on the
\textsc{Helac-Phegas} leading-order event generator for all parton level 
processes \cite{Kanaki:2000ey,Papadopoulos:2000tt,Cafarella:2007pc}. 
The NLO system consists of:
1) \textsc{CutTools} \cite{Ossola:2007ax}, for the OPP reduction 
of tensor integrals with a given numerator to a basis of scalar functions
and for the rational parts 
\cite{Ossola:2006us,Ossola:2008xq,Draggiotis:2009yb}; 
2) \textsc{Helac-1Loop} \cite{vanHameren:2009dr} for the evaluation 
of one loop amplitude, more specifically for the evaluation of 
the numerator functions for given loop momentum (fixed by \textsc{CutTools}); 
3)  \textsc{OneLOop} \cite{vanHameren:2009dr}, 
a library of scalar functions, which provides the actual numerical values of 
the integrals. 
4) \textsc{Helac-Dipoles} \cite{Czakon:2009ss}, automatic implementation  of 
Catani-Seymour dipole subtraction \cite{Catani:2002hc}, for the 
calculation of the real emission part.

Let us emphasize that all parts are calculated fully numerically  
in a completely automatic manner. 

\section{Numerical results}

Proton-proton collisions at the LHC with a center of mass energy
of $\sqrt{s}=14$ TeV are considered.  The mass of the top quark is set  
to be $m_t=172.6$ GeV. We leave it on-shell with unrestricted 
kinematics. The jets are defined by at
most  two partons using the $k_T$ algorithm with a separation $\Delta R=0.8$,
where $\Delta R=\sqrt{(y_1-y_2)^2+(\phi_1-\phi_2)^2}$,
$y_i=1/2\ln(E_i-p_{i,z})/(E_i+p_{i,z})$ being the rapidity and $\phi_i$ the
azimuthal angle of parton $i$. Moreover, the recombination is only performed
if both partons satisfy $|y_i|<5$ (approximate detector bounds). We further
assume for $t\bar{t}b\bar{b}$  ($t\bar{t}jj$) processes, that the jets are
separated by  $\Delta R=0.8$ $(1.0)$ and have $|y_{\rm{jet}}| < 2.5$ $(4.5)$.  
Their transverse momentum is required to be larger than $20$ $(50)$ GeV 
respectively.  We consistently use the CTEQ6 set of parton distribution
functions, i.e.  we take CTEQ6L1 PDFs with a 1-loop running $\alpha_s$ in LO
and CTEQ6M PDFs with a 2-loop running $\alpha_s$ at NLO.
%------------------------------------------------------------------------------%
\begin{figure}
{\begin{center}
\psfig{figure=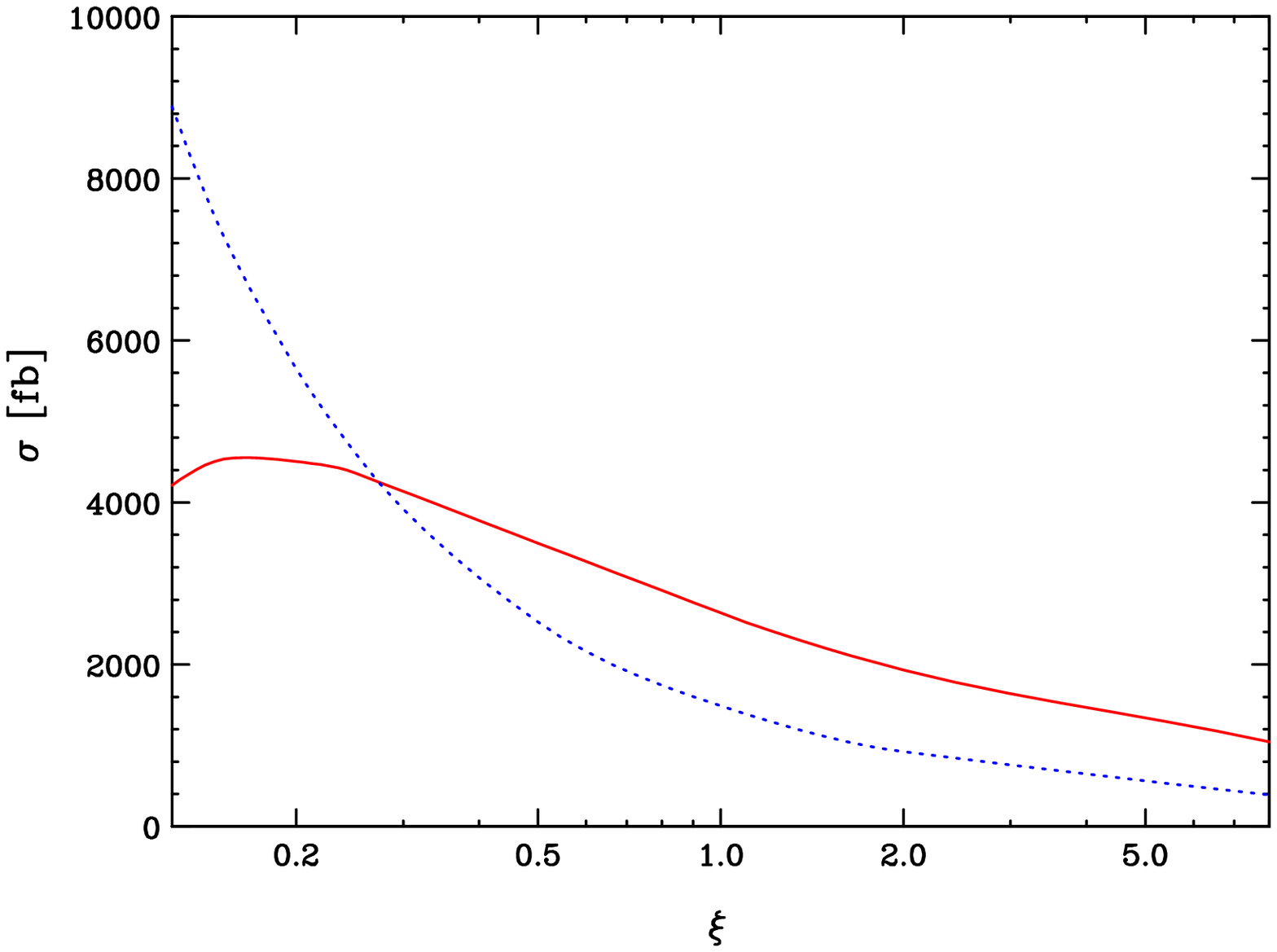,height=2.1in}
\psfig{figure=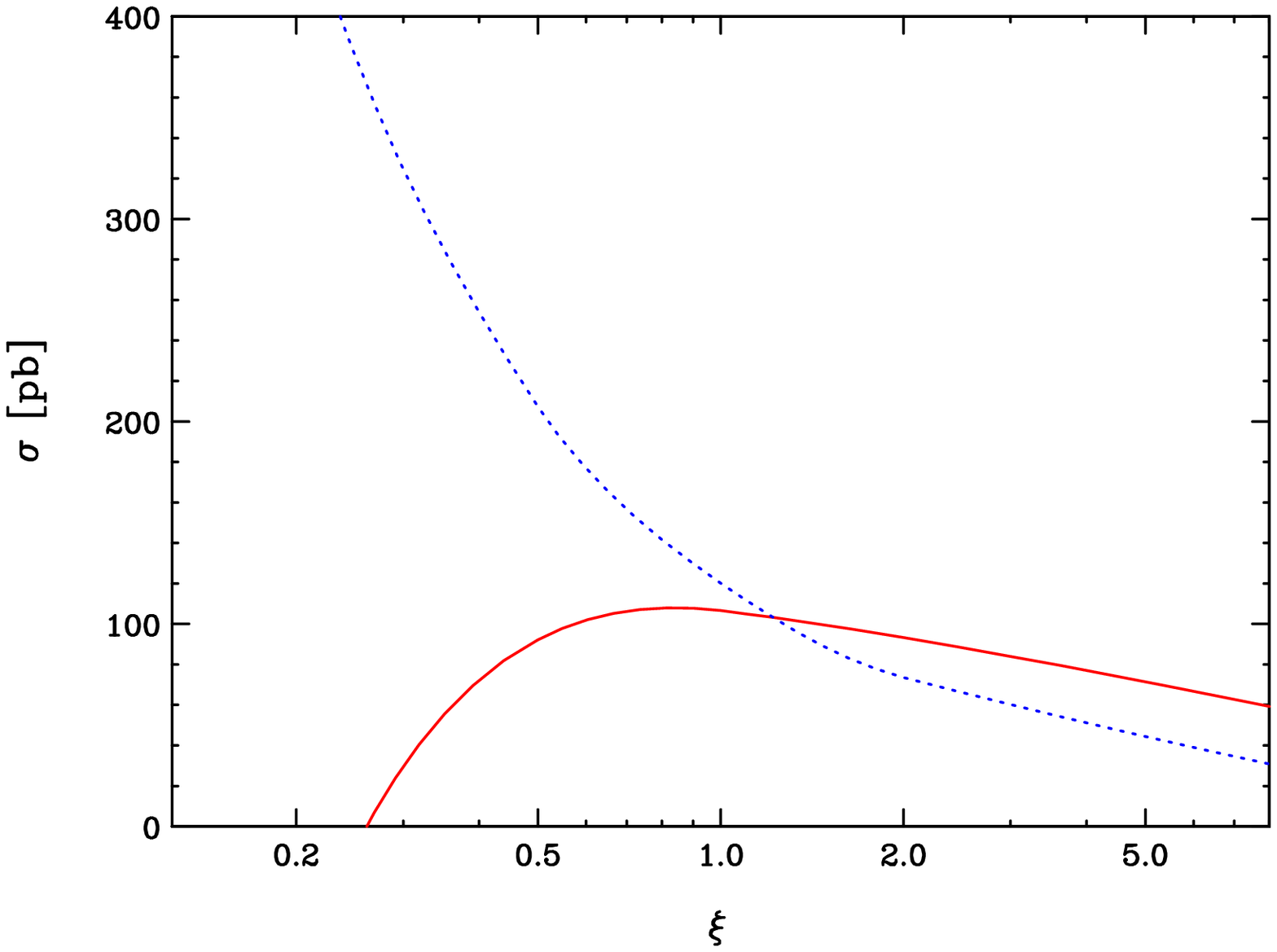,height=2.1in}
\end{center}}
\caption{Scale dependence of the total cross section for $pp\rightarrow
  t\bar{t}b\bar{b} + X$ (left panel) and for $pp\rightarrow t\bar{t} jj + X$
  (right panel) at the LHC   with $\mu_R=\mu_F=\xi \cdot \mu_0$ where
  $\mu_0=m_t=172.6$ GeV.   The blue dotted curve corresponds to  the LO
  whereas the red solid to the NLO one.
\label{fig:scales}}
\end{figure}
%------------------------------------------------------------------------------%

We begin our presentation of the final results of our analysis with a
discussion of the total cross section. At the central value of the
scale, $\mu_R=\mu_F=\mu_0=m_t$, we have obtained
\[
\sigma^{\rm{LO}}_{pp\rightarrow t\bar{t}b\bar{b}+X}= 
{1489.2 ~}^{\rm{+1036.8 (70\%)}}
_{-\rm{~565.8 (38\%)}} ~~\mbox{fb}\;,
~~~~~~\sigma^{\rm{NLO}}_{pp\rightarrow t\bar{t}b\bar{b}+X}= 
{2636 ~}^{\rm{+862 (33\%)}}
_{-\rm{703 (27\%)}} ~~\mbox{fb} \;,
\]
\[
\sigma^{\rm{LO}}_{pp\rightarrow t\bar tjj+X} = {120.17 ~}^{\rm{+ 87.14(72\%)}}
_{\rm{- 46.64 (39\%)}}  ~~\rm{pb}\;,
~~~~~~
\sigma^{\rm{NLO}}_{pp\rightarrow t\bar tjj+X} = {106.94 ~}^{\rm{- 14.30(13\%)}}
_{\rm{- 13.28 (12\%)}}  ~  ~\rm{pb}\;,
\]
where the error comes from varying the scale up and down by a factor 2. From
the above result one can obtain $K$ factors
\[
K_{pp\rightarrow t\bar{t}b\bar{b}+X}= 1.77\;,
~~~~~~K_{pp\rightarrow t\bar tjj+X} =  0.89\;.
\]
In case of $pp\rightarrow t\bar{t}b\bar{b}+X$ corrections are large of the
order of  $77\%$.  However, they can be reduced substantially, even down to
$-11\%$,   either by applying   additional cuts or by a better choice of 
factorization and renormalization  scales as already suggested by Bredenstein
et al. \cite{Bredenstein:2010rs}.   In case of $pp\rightarrow t\bar{t}jj+X$ we
have obtained negative  corrections of the order of 11\%. In both cases a
dramatic reduction of the scale uncertainty is observed while going from LO to
NLO. The scale dependence of the corrections for both processes is graphically 
presented in Fig.~\ref{fig:scales}.

\begin{figure}
{\begin{center}
\psfig{figure=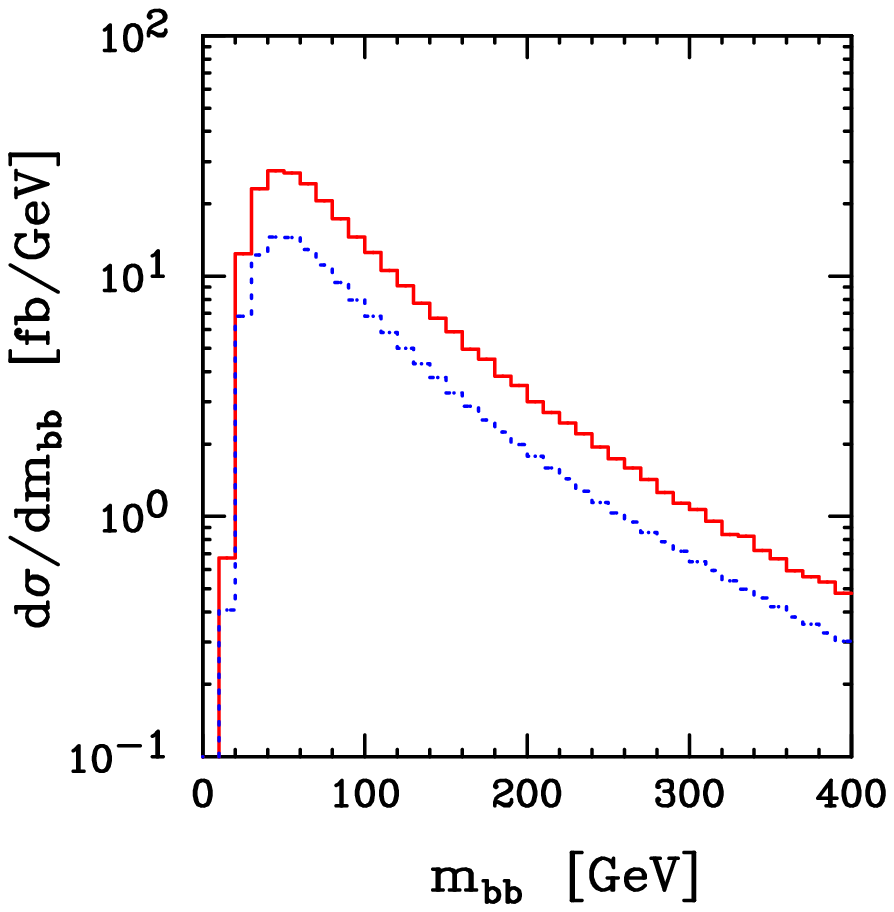,height=2.5in}
\psfig{figure=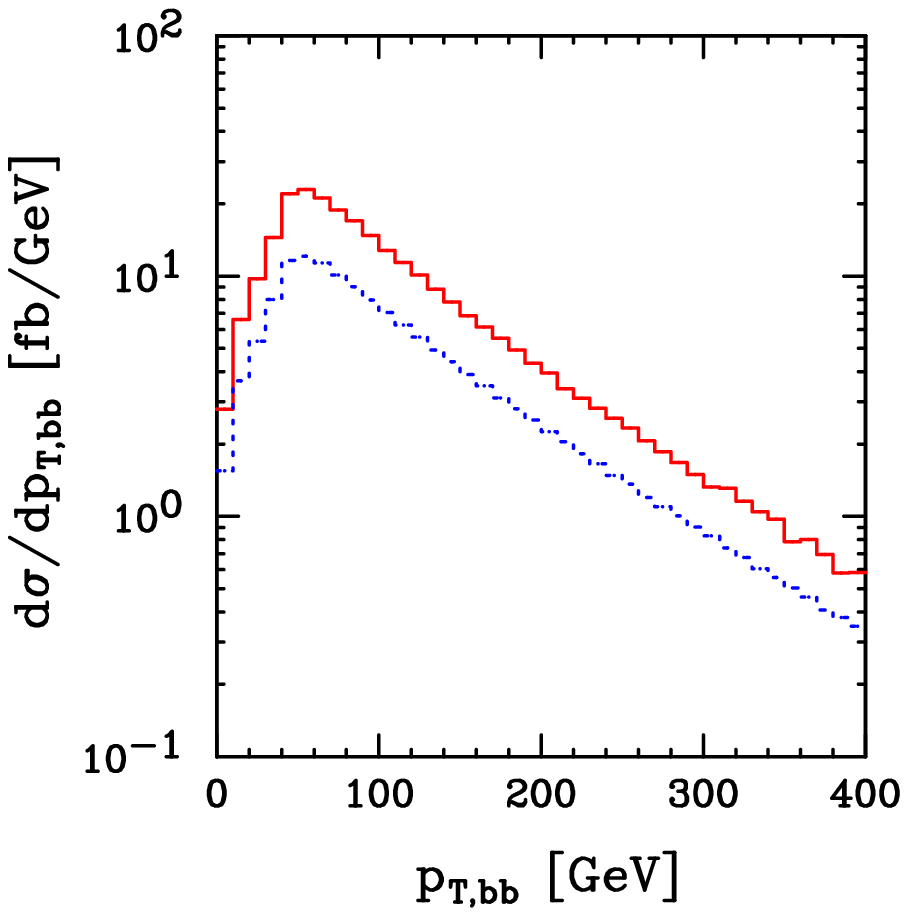,height=2.5in}
\end{center}}
\caption{Distribution of the invariant mass $m_{b\bar{b}}$ (left panel) and
  the distribution in the transverse momentum  $p_{T_{b\bar{b}}}$  (right
  panel)  of the bottom-anti-bottom  pair for $pp\rightarrow t\bar{t}b\bar{b}
  + X$ at the LHC.  The blue dotted curve corresponds to  the LO  whereas the
  red solid to the NLO one.
\label{fig:ttbb}}
\end{figure}

While the size of the corrections to the total cross section is
certainly interesting, it is crucial to study the corrections to the
distributions.
In Fig.~\ref{fig:ttbb} the differential distributions for two
observables, namely the invariant mass and transverse momentum
of the two-$b$-jet system are depicted for the $pp\rightarrow t\bar{t}b\bar{b}
  + X$ process. Clearly, the distributions show
the same large corrections, which turn out to be relatively constant
contrary to the quark induced case \cite{Bredenstein:2008zb}.
%------------------------------------------------------------------------------%
%------------------------------------------------------------------------------%
\begin{figure}
{\begin{center}
\psfig{figure=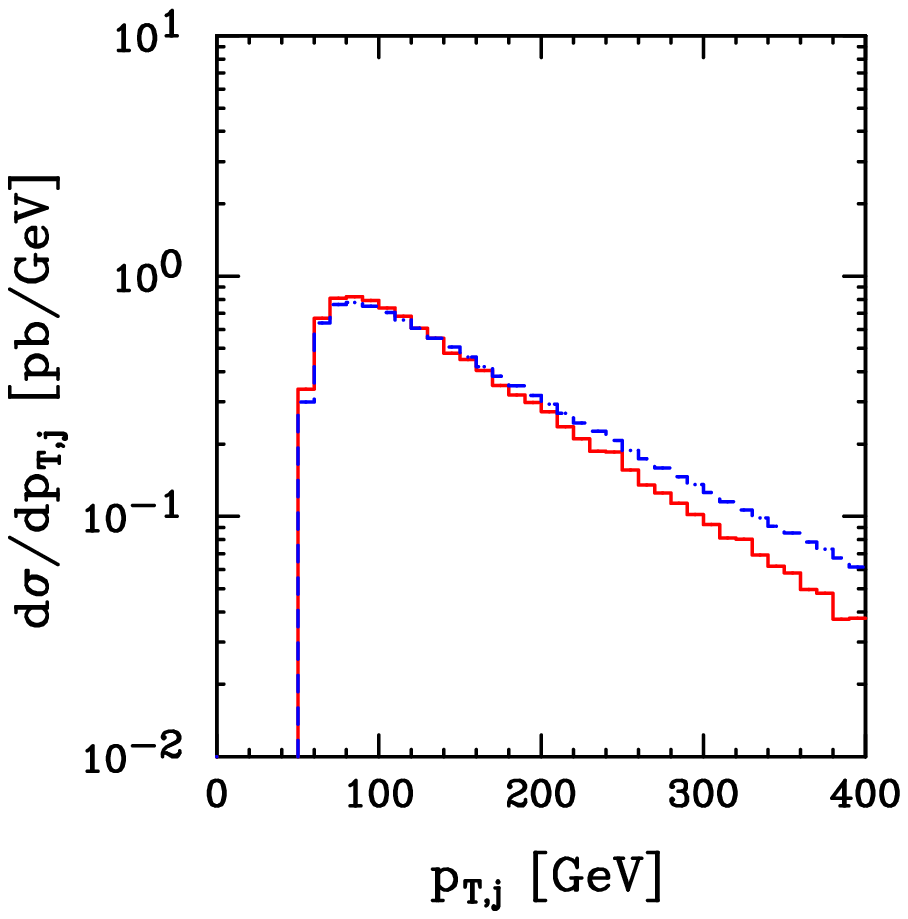,height=2.5in}
\psfig{figure=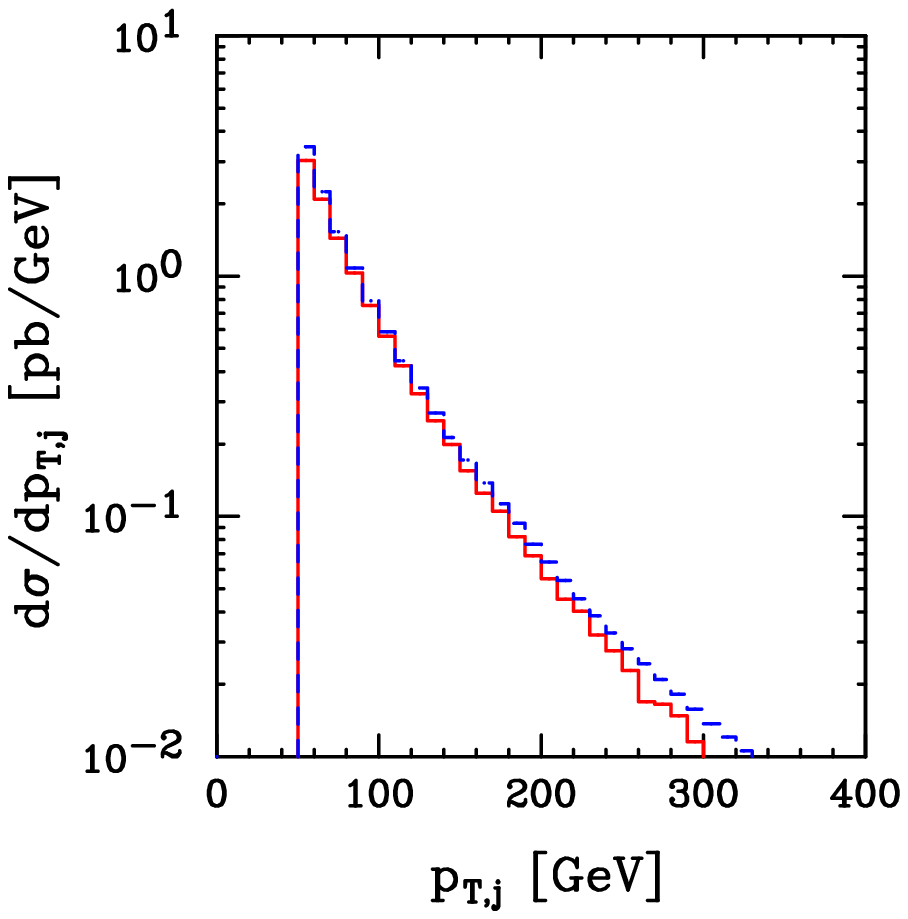,height=2.5in}
\end{center}}
\caption{Distribution in the transverse momentum $p_{T_{j}}$  of the   1st
  hardest jet (left panel) and the 2nd hardest jet (right panel)  for
  $pp\rightarrow t\bar{t} jj +X$ at the LHC.  The blue dotted curve
  corresponds to  the LO  whereas the red solid to the NLO one.
\label{fig:ttjj}}
\end{figure}
%------------------------------------------------------------------------------%
In Fig.~\ref{fig:ttjj} the transverse momentum distributions of the hardest 
and  second hardest jet are shown for the $pp\rightarrow t\bar{t}jj+ X$ 
process. Distributions demonstrate tiny corrections
up to at least 200 GeV, which means that the size of the corrections
to the cross section is transmitted to the distributions.
On the other hand, strongly altered shapes are  visible
at high $p_T$ especially in case of the first hardest jet. Let us underline
here, that corrections to the high $p_T$ region can only be correctly 
described by higher order calculations and are not altered by soft-collinear 
emissions simulated by parton showers.

\section*{Conclusions}

A brief summary of the calculations of NLO QCD corrections to the background
processes $pp\rightarrow t\bar{t}b\bar{b} + X$ and  $pp\rightarrow t\bar{t}jj
+ X$ at the LHC has been presented. They have been calculated 
with the help of the \textsc{Helac-Nlo} system. 

The QCD corrections to the integrated
cross  section for the irreducible background   are found to be very large,
changing the LO results by about 77\%.  The distributions show the same large
corrections which are relatively constant.  
The residual scale uncertainties of the NLO predictions are at
the 33\% level.  
On the other hand, the corrections to the reducible background with respect to
LO are negative and small, reaching 11\%. The error obtained by scale
variation is  of the same order.  The size of the corrections to the cross
section is transmitted to the  distributions at least for the low $p_T$ region. 
However, the shapes change appreciably  at high $p_T$.

\section*{Acknowledgments}

I would like to thank the organizers of “Recontres de Moriond” for  the kind
invitation and the very pleasant atmosphere during the conference.  The work
presented here was  funded by the Initiative and Networking Fund of the
Helmholtz Association,  contract HA-101 (``Physics at the Terascale'') and by
the RTN European Programme MRTN-CT-2006-035505  \textsc{Heptools} - Tools and
Precision Calculations for  Physics Discoveries at Colliders.

\end{document}